\documentclass{article}
\usepackage[T1]{fontenc}
\usepackage{graphics}

\makeatletter

\providecommand{\LyX}{L\kern-.1667em\lower.25em\hbox{Y}\kern-.125emX\@}

 \newcommand{\lyxaddress}[1]{
   \par {\raggedright #1 
   \vspace{1.4em}
   \noindent\par}
 }

\makeatother

\begin{document}

\title{Effect of a rigid nonpolar solute on the splay, bend elastic constants and
on rotational viscosity coefficient of \( 4,4^{\prime } \) -n-octyl-cyanobiphenyl}

\author{Sudeshna DasGupta, Pinku Chattopadhyay and Soumen Kumar Roy}

\date{\today}

\maketitle

\lyxaddress{Department of Physics, Jadavpur University, Calcutta-700 032, INDIA }

\begin{abstract}
The effect of a rigid nonpolar non-mesogenic solute, {}``biphenyl{}'' which
is \( C_{6}H_{5}-C_{6}H_{5} \), on the splay and bend elastic constants and
on the rotational viscosity coefficient of \( 4,4^{\prime } \)-n-octyl-cyano
biphenyl (8CB) is reported. The experiments involve the measurement of voltage
dependence of capacitance of a cell filled with the mixture. Anomalous behavior
of both \( K_{11} \) and \( \Delta \epsilon  \) near the \( N-S_{A} \) transition
have been observed. 
\end{abstract}

\section{INTRODUCTION}

It is well known that, with few exceptions, the presence of non-mesomorphic
solutes in nematics depress the normal nematic-isotropic transition temperature
(\( T_{NI} \)) \cite{martire76}. Moreover, experiments \cite{peterson74,kroberg76}
over last few decades have confirmed that the presence of such impurities leads
to the formation of a two phase region. This is consistent with the laws of
thermodynamics and the first order nature of the \( N-I \) transition. Two
types of models have proved to be generally useful for investigating \( N-I \)
transitions in dilute solutions. The first is a lattice model \cite{cotter69,cotter76}
which takes into account the role of intermolecular repulsion or the steric
effect. The other \cite{maier59} is the mean field theory of the Maier-Saupe
type which takes into account the anisotropic intermolecular attraction or dispersion
forces. More recently Mukherjee \cite{mukherjee97} has explained the behavior
of the two phase region using the phenomenological Landau de Gennes theory.

The role of non mesogenic impurities on a mesogen which exhibits a smectic phase
as well, has not received a great deal of attention of either experimentalists
or theorists. Here too, in favourable circumstances, remains a possibility of
the existence of a two phase region near the smectic transition \cite{deGennesch2}. 

In this paper we report the results of measurements carried out on a mixture
of biphenyl and \( 4,4^{\prime } \)-n-octyl-cyano biphenyl (8CB). The latter
which has a strong polar group and a long alkyl chain has received considerable
attention of experimentalists over last few decades \cite{karat77,bradshaw85,morris86,pinku94}
and is known to exhibit a first order \( S_{A}-N \) transition. The biphenyl+8CB
system to our knowledge, has not been studied before and is a natural and interesting
choice in its own merit since the solute, which is chemically \( C_{6}H_{5}-C_{6}H_{5} \),
is a non-mesogenic, non-polar, rigid molecule and is nothing but 8CB deprived
of its cyano group and the flexible alkyl chain. We have observed that biphenyl
is easily miscible with 8CB and lowers both \( T_{AN} \) and \( T_{NI} \)
by an amount which depends on the impurity concentration. The range of the nematic
phase i.e., \( (T_{NI}-T_{AN}) \) however is not significantly altered. The
effect of this rod-like non-polar impurity on the dielectric anisotropy, the
Frank elastic constants (splay and bend) and on the effective rotational viscosity
coefficient of pure 8CB has been reported here. To our knowledge, no measurement
has yet been reported on the effect of impurities on the elastic constants and
the viscosity coefficient of a nematic. The experiments we have carried out
involve electric field induced Freedericksz transition and consist of two parts.
The dielectric anisotropy and the Frank elastic constants are extracted from
a set of measurements of the voltage dependence of capacitance of a sample cell
filled with the 8CB and biphenyl mixture. For the determination of the rotational
viscosity coefficient a method based on dynamic Freedericksz transition is used.
Interesting features have emerged from these measurements and the splay elastic
constant is seen to exhibit anomalous behavior in the vicinity of the \( S_{A}-N \)
transitions for the impurity concentration exceeding a certain value.

\section{PRINCIPLE OF THE METHOD}

The splay elastic constant \( K_{11} \) is related to the Freedericksz threshold
voltage \( V_{th} \) by the relation

\begin{equation}
\label{k11eq}
K_{11}=\epsilon _{0}\Delta \epsilon /\pi ^{2}V_{th}^{2}
\end{equation}
 where \( \epsilon  \)\( _{0} \) is the permittivity of free space and \( \Delta  \)\( \epsilon =\epsilon _{\Vert }-\epsilon _{\bot } \),
the dielectric anisotropy of the sample. The method of determination of \( \Delta \epsilon  \)
is described below.

The exact relationship between the cell capacitance \( C \) and the voltage
\( V \) applied across the cell was obtained by Gruler \cite{gruler}

\begin{equation}
\label{vrat}
\frac{V}{V_{th}}=\frac{2}{\pi }(1+\gamma \sin ^{2}\phi _{m})^{1/2}\int _{0}^{\phi _{m}}\left[ \frac{(1+\kappa \sin ^{2}\phi )}{(1+\gamma \sin ^{2}\phi )(\sin ^{2}\phi _{m}-\sin ^{2}\phi )}\right] ^{1/2}d\phi 
\end{equation}
 and

\begin{equation}
\label{crat}
\frac{C}{C_{\bot }}=\frac{\int _{0}^{\phi _{m}}\left[ \frac{(1+\kappa \sin ^{2}\phi )(1+\gamma \sin ^{2}\phi )}{(\sin ^{2}\phi _{m}-\sin ^{2}\phi )}\right] ^{1/2}d\phi }{\int _{0}^{\phi _{m}}\left[ \frac{(1+\kappa \sin ^{2}\phi )}{(1+\gamma \sin ^{2}\phi )(\sin ^{2}\phi _{m}-\sin ^{2}\phi )}\right] ^{1/2}d\phi }
\end{equation}
 where \( \kappa =K_{33}/K_{11}-1 \) (\( K_{33} \) being the bend elastic
constant), \( \gamma =\epsilon _{\Vert }/\epsilon _{\bot }-1 \), \( \phi  \)
is the tilt angle made by the director with a direction parallel to the cell
walls, \( \phi  \)\( _{m} \) is the tilt angle at the center of the cell and
\( C_{\bot } \) is the capacitance of the cell when the liquid crystal molecules
are homogeneously aligned, that is, before the onset of Freedericksz transition
when the voltage applied is lower than the threshold voltage. These equations
can be combined to yield

\begin{equation}
\label{crat1}
\frac{C-C_{\bot }}{C_{\bot }}=\gamma -\frac{2\gamma }{\pi }(1+\gamma \sin ^{2}\phi _{m})^{1/2}\frac{V_{th}}{V}\int _{0}^{\sin \phi _{m}}\left[ \frac{(1+\kappa x^{2})(1-x^{2})}{(1+\gamma x^{2})(\sin ^{2}\phi _{m}-x^{2})}\right] ^{1/2}dx
\end{equation}
 When the applied voltage is much higher that the threshold voltage, the director
at the center of the cell becomes perpendicular to the cell walls and \( \phi _{m}=\pi /2 \).
Then the above equation reduces to 

\begin{equation}
\label{crat2}
\frac{C-C_{\bot }}{C_{\bot }}=\gamma -\frac{2\gamma }{\pi }(1+\gamma )^{1/2}\frac{V_{th}}{V}\int _{0}^{1}\left[ \frac{(1+\kappa x^{2})}{(1+\gamma x^{2})}\right] ^{1/2}dx
\end{equation}
 or dividing by \( \gamma  \)

\begin{equation}
\label{cred}
\frac{C-C_{\bot }}{C_{\Vert }-C_{\bot }}=C_{R}=1-\frac{2}{\pi }(1+\gamma )^{1/2}\frac{V_{th}}{V}\int _{0}^{1}\left[ \frac{(1+\kappa x^{2})}{(1+\gamma x^{2})}\right] ^{1/2}dx
\end{equation}
 where \( C_{R} \) may be called the reduced capacitance. \( C \)\( _{\Vert } \)
is the capacitance of the cell when the nematic is homeotropically oriented,
i.e., the value of \( C \) in the limit \( 1/V \) \( \rightarrow 0 \). \( \epsilon _{\Vert } \)
and \( \epsilon _{\bot } \) may be obtained by dividing \( C_{\Vert } \) and
\( C_{\bot } \) by the empty cell capacitance \( C_{0} \) respectively. In
the above equation use has been made of the relation \( \gamma =(C_{\Vert }-C_{\bot })/C_{\bot } \). 

Thus equation (\ref{crat2}) predicts that a plot of \( (C-C \)\( _{\bot })/C_{\bot } \)
against \( 1/V \) for \( V \)\( \gg V_{th} \) should be linear and the extrapolated
value of the ordinate for \( 1/V \) \( \rightarrow 0 \) should directly provide
the value of \( \gamma  \)=\( \Delta \epsilon /\epsilon _{\bot } \). This
procedure for obtaining \( \gamma  \) was first suggested by Meyerhofer \cite{meyerhofer}. 

The variation of capacitance with applied voltage was fitted against equations
(\ref{vrat}) and (\ref{crat}) to obtain \( V_{th} \) and \( \kappa  \).
We then used (\ref{k11eq}) to calculate \( K_{11} \) and hence \( K_{33} \).
This method for obtaining \( V_{th} \) and \( \kappa  \) was suggested by
Morris et al \cite{morris86}. 

The viscous behavior of a nematic can be obtained by studying the response to
a sudden change of an applied magnetic or electric field which is normally stronger
than the Freedericksz-threshold value. In case of deformations involving a pure
twist there is no hydrodynamic flow. The molecules merely rotate without any
translational motion and the analysis is rather simple \cite{dejeu}. The situation
is far more complicated in case of a splay geometry which we have studied since
the director reorientation is now accompanied by a hydrodynamic flow. The gradient
of the angular velocity of the director produces a backflow motion, first demonstrated
by Pieranski \cite{pieranski73} giving rise to a frictional torque. The sudden
removal of an external field in this case would result in a director relaxation
time

\begin{equation}
\label{rotvisc}
\tau =\gamma _{1}^{*}d^{2}/(\pi ^{2}K_{11})
\end{equation}
 where \( \gamma _{1}^{*} \) is the effective rotational viscosity coefficient,
\( d \) is the cell thickness and \( K_{11} \) is the splay elastic constant.
In practice \( \gamma _{1}^{*} \) is about \( 10\% \) less than the rotational
viscosity coefficient \( \gamma _{1} \) .

We have determined the relaxation time \( \tau  \) by measuring the decay of
capacitance in a nematic cell. The details are given in the next section.

\section{DETAILS OF THE EXPERIMENTAL METHOD}

We have used a Hewlett-Packard LCR meter HP 4274A to measure the capacitance
of the sample cell. The instrument employs the so called {}``auto balancing
bridge method{}'' to measure both the real and complex parts of impedance simultaneously.
The voltage was varied from 0.1 V to 5 V at a frequency of 1KHz. The sample
cells, which were of very high precision, were obtained from Displaytech, USA.
The cells consisting of two ITO coated glass plates with a spacing of 4 \( \mu  \)m,
and an active area of 0.26 cm\( ^{2} \), had brushed polyimide treatment to
ensure planar orientation. They also had a guard ring incorporated in them to
minimize fringe electric fields. The temperature of the cells were controlled
to within \( \pm 0.1K \) by placing the cells within a Mettler hot stage FP
82. The texture of the samples were viewed through a Leitz polarizing microscope
and photographs were taken with the help of Photoautomat. Both 8CB and biphenyl
were obtained from Merck and were used without further purification.

\begin{figure}
{\par\centering \resizebox*{0.9\textwidth}{!}{\rotatebox{270}{\includegraphics{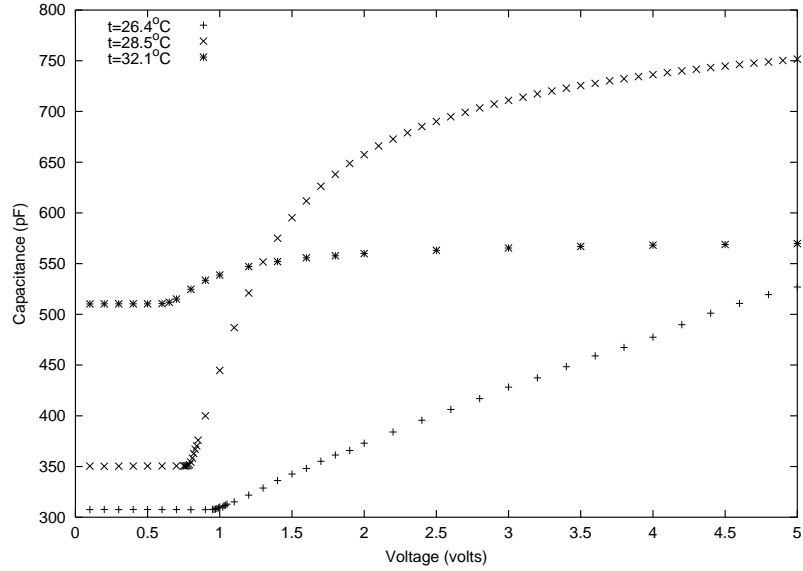}}} \par}

\caption{\label{c-v variation}Variation of capacitance \protect\( C\protect \) with
voltage \protect\( V\protect \) for different temperatures \protect\( t\protect \)
for a \protect\( 4.59\%\protect \) mixture of biphenyl in 8CB.}
\end{figure}

The samples which we have used were mixtures of various concentrations of biphenyl,
ranging from 0.4\% to 6.4\%, in pure 8CB. The smectic to nematic transition
temperature, \( T_{AN} \) and the nematic to isotropic transition temperature
\( T_{NI} \) were determined from texture studies of the samples. Various photographs
were taken at different stages of transition. In the capacitance measurement
experiments the probe voltage of the LCR meter HP 4274A provided the aligning
a.c. electric field. The voltage was varied from 0.1 V to 5 V at a frequency
of 1KHz in the \( C_{P}-G \) mode of the instrument. In the neighborhood of
Freedericksz transition readings were taken at intervals of 10mV. The samples
were filled in the cells and the C-V variation was recorded at a large number
of temperatures ranging from the smectic to nematic transition temperature,
\( T_{AN} \), to a few degrees beyond the nematic to isotropic transition temperature,
\( T_{NI} \). It must be noted that all measurements were carried out as the
temperatures of the sample were increased gradually starting from \( T<T_{AN} \)
and none of the measurements were taken during cooling. Fig.\ref{c-v variation}
shows a few capacitance-voltage variations of the mixture with biphenyl concentration
\( 4.59\% \) at temperatures near the two transitions \( S_{A}-N \) and \( N-I \)
and at an intermediate temperature. There is always a need to have a refinement
in the value of \( V_{th} \) as the C-V curves obtained in a polyimide cell
usually do not exhibit a very sharp threshold of Freedericksz transition so
as to yield a precise value of \( V_{th} \) from the data directly. From the
C-V data \( \epsilon _{\bot } \) and \( \epsilon _{\Vert } \) were calculated
and subsequently \( V_{th} \) and \( \kappa  \) were determined as described
below. \( \epsilon _{\bot } \) was determined directly from the value of capacitance
at voltages lower than the threshold voltage, (before the onset of the Freedericksz
transition) by dividing the capacitance, \( C_{\bot } \) by the empty cell
capacitance \( C_{0} \). \( \epsilon _{\Vert } \) was obtained by plotting
\( C \) against \( 1/V \) for the higher voltage range \( (4.0-5.0V) \).
The extrapolated value of capacitance in the limit as \( 1/V\rightarrow 0 \)
gives \( C_{\Vert } \) which when divided by \( C_{0} \) gives \( \epsilon _{\Vert } \)
. 

A direct measurement of \( C_{\Vert } \) was also carried out to check the
reliability of the value of \( \gamma  \) obtained by the above method. A homeotropic
alignment was obtained in the ITO coated glass cells using CTAB (cetyl trimethyl
ammonium bromide) and the capacitance of the cell was measured while a \( 10KG \)
magnetic field was applied across the cell. The values of \( \gamma  \) thus
obtained never differed from the \( \gamma  \) 's obtained from the \( 1/V \)
-extrapolation method by more than \( 0.1\% \). 

A two parameter non-linear least square fit for finding \( V_{th} \) and \( \kappa  \)
worked in the following manner. Starting values of \( V_{th} \) and \( \gamma  \)
were taken directly from the experiment. (\ref{vrat}) was then used to obtain
the \( \phi _{m} \) 's for all values of V for which measurements of capacitance
were carried out. The \( \phi _{m} \) 's thus obtained were used in (\ref{crat})
to obtain C. The error which was minimized in the least square programme is
\( \sum _{i=1}^{n}(C_{i}^{expt}-C_{i})^{2} \) where \( C_{i}^{expt} \) is
the measured value of the capacitance at V=\( V_{i} \) and \( C_{i} \) is
the value of capacitance obtained by solving (\ref{crat}), n being the number
of data points for a particular temperature. The final values of \( V_{th} \)
never differed a great deal from the input values.

For the rotational viscosity coefficient measurements, we have used a Hewlett-Packard
impedance gain-phase analyser HP \( 4194 \)A in the programmable mode to record
the decay of capacitance with time. An integration time of \( 500 \) \( \mu  \)sec
was chosen which resulted in an interval of \( 5 \) msec between the successive
readings at the operating frequency which was fixed at \( 10 \) KHz. With our
instrument a delay time, in multiples of \( 1 \) msec, could be introduced
between the readings but in this case we did not introduce any delay since the
decay of capacitance was found to be pretty fast. The a.c. probe voltage across
the sample cell was held fixed at \( 0.3 \) V. A low frequency a.c. voltage
about \( 2 \) volts peak to peak was applied across the cell for about \( 30 \)
secs. It was switched off and the transient (decaying) capacitance of the cell,
was captured as a function of time by the HP \( 4194 \)A. A plot of this decay
of capacitance with time which was always found to be exponential was then dumped
into a HP \( 7475 \)A plotter which was connected to the impedance analyser
over a HPIB bus. From the plots thus obtained for different concentration of
biphenyl mixtures, at different temperatures, the time constant, \( \tau  \)
was calculated and using the values of \( K_{11} \) obtained from the static
capacitance measurements, the effective rotational viscosity coefficient, \( \gamma _{1}^{*} \)
was evaluated.

\section{RESULTS AND DISCUSSIONS}

\begin{figure}
{\par\centering \resizebox*{0.9\textwidth}{!}{\rotatebox{270}{\includegraphics{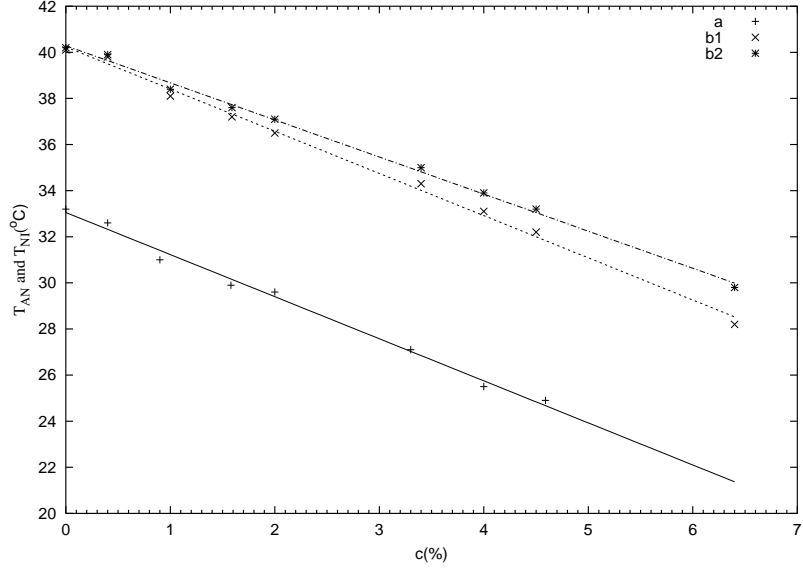}}} \par}

\caption{\label{phase diagram}Variation of \protect\( T_{NI}\protect \) and \protect\( T_{AN}\protect \)
with concentration of biphenyl (c) . (a =\protect\( T_{AN}\protect \) , b1=\protect\( T_{NI}\protect \)
(lower), b2=\protect\( T_{NI}\protect \) (upper))}
\end{figure}

The transition temperatures \( T_{NI} \) (upper), \( T_{NI} \) (lower) and
\( T_{AN} \) were determined from optical texture studies and the results are
presented in Table \ref{transition temp(table)} and Fig.\ref{phase diagram}.
The \( T_{NI} \) (upper) is the temperature at which the field of view of the
polarizing microscope (with crossed polarizer and analyzer) becomes completely
dark. All real nematics contain impurities and Rosenblatt \cite{rosenblatt82}
has observed in 8CB a two phase \( (N+I) \) region of temperature of width
\( 13mK \) in absence of an electric field. Needless to say, with our temperature
resolution of \( 0.1K \), we were unable to confirm this. The width of the
two phase region was seen to increase with the concentration of biphenyl and
for the maximum concentration of \( 6.4\% \) this was \( 1.6K \). We were
unable to work with higher impurity concentration because of non-availability
of an experimental arrangement for producing sub-ambient temperatures. All the
transition temperatures reported in Table \ref{transition temp(table)} were
determined without the presence of any external field. However, we confirmed
that with the highest electric field we applied across the sample cells, namely
\( 1.25\times 10^{4} \) \( volt/cm \), there was no noticeable change in any
of the transition temperatures.

\begin{figure}
{\par\centering \resizebox*{0.9\textwidth}{!}{\includegraphics{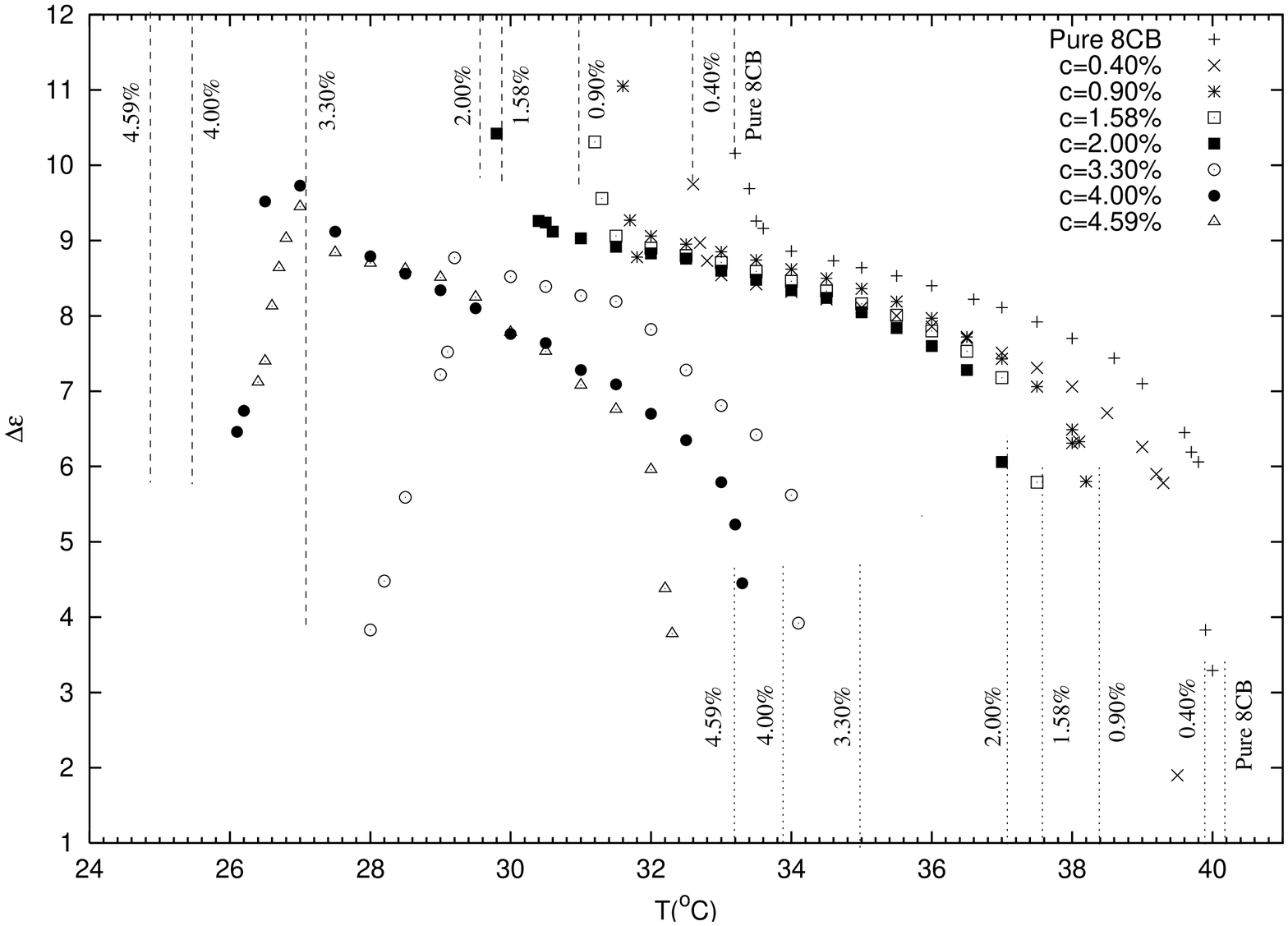}} \par}

\caption{\label{deltaeps}Variation of \protect\( \Delta \protect \)\protect\( \epsilon \protect \)
with \protect\( T\protect \) for different concentrations of biphenyl in 8CB.
The vertical lines denote the transition temperatures \protect\( T_{AN}\protect \)
and \protect\( T_{NI}\protect \) for different concentrations of biphenyl in
8CB as shown in Table \ref{transition temp(table)}.}
\end{figure}

The temperature dependence of the dielectric anisotropy for the concentration
(c) of biphenyl ranging from \( 0 \) to \( 4.59\% \) is shown in Fig.\ref{deltaeps}.
In pure 8CB and for all mixtures we have studied \( \Delta \epsilon  \), in
the neighborhood of \( T_{NI} \), shows the same feature in that it disappears
at \( T_{NI} \) (upper). At this particular temperature the field of view in
the microscope with crossed polars is completely dark and hence the phase is
isotropic. Values of the dielectric constant \( \epsilon _{iso} \) obtained
well within the isotropic phase show a gradual increase with temperature for
pure 8CB as well as for all mixtures we have studied. 

Focussing on the behavior of \( \Delta \epsilon  \) as \( T_{AN} \) is approached
from the higher temperature side we find that for biphenyl concentration, c
upto 2.00\% \( \Delta \epsilon  \) in mixtures behaves the same way as that
in pure 8CB, namely that it shows a sharp increase. However, for impurity concentrations
of \( 3.30\% \) and higher, \( \Delta \epsilon  \) goes down sharply, while
always remaining positive (at least in the temperature range we have explored).
We have noted that the decrease in \( \Delta \epsilon  \) results from a reduction
of \( \epsilon _{\Vert } \), (\( \epsilon _{\bot } \) remaining fairly constant)
as \( T_{AN} \) is approached. A similar decrease in \( \epsilon _{\Vert } \)
as \( T_{AN} \) is approached has been observed in pure \( p,p^{\prime } \)
- diheptylazoxybenzene where the \( N-S_{A} \) transition is almost second
order. This has been attributed to the presence of pretransitional effects within
the nematic phase \cite{deJeu74}. 

\begin{figure}
{\par\centering \resizebox*{0.9\textwidth}{!}{\includegraphics{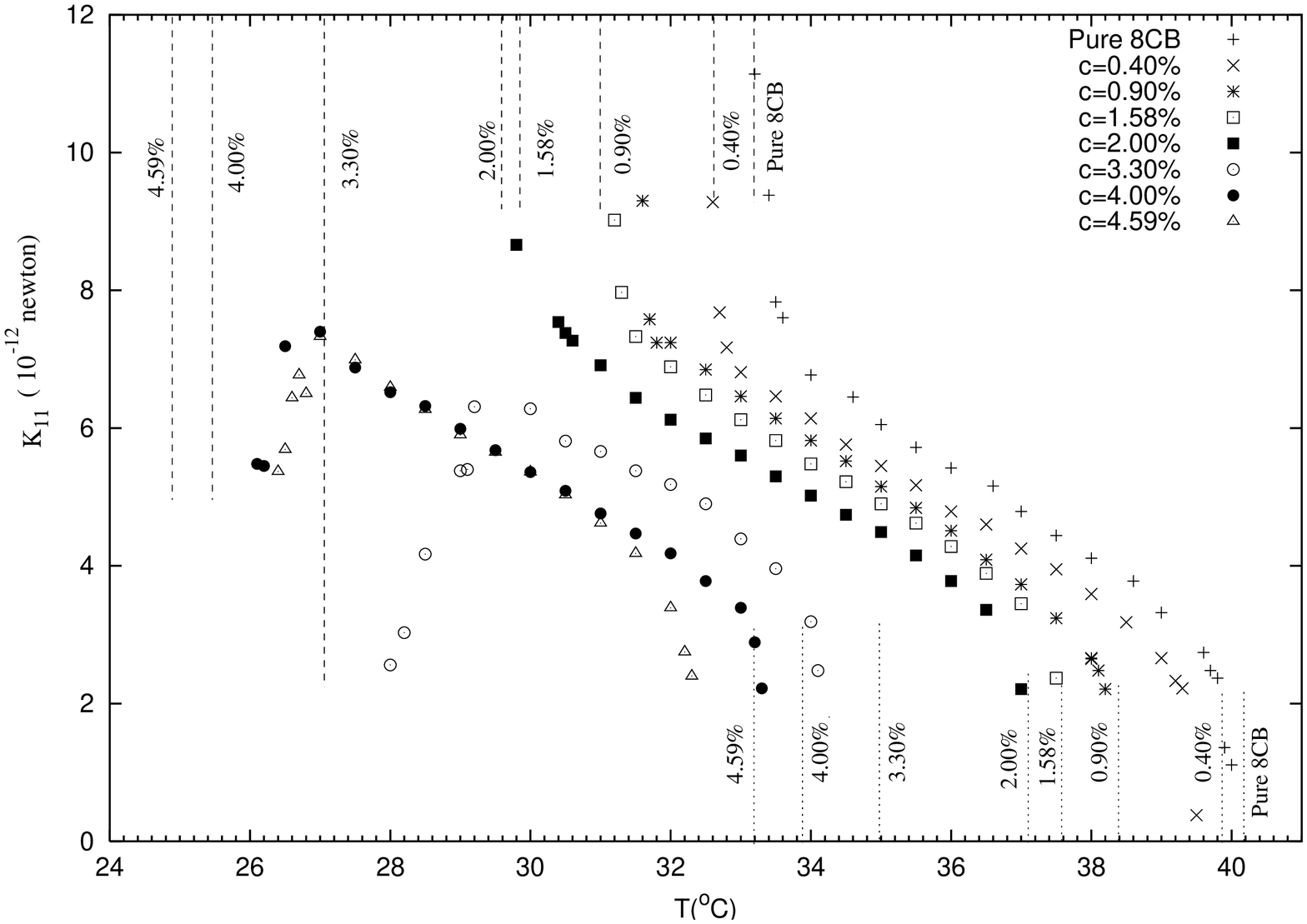}} \par}

\caption{\label{k11}Variation of \protect\( K_{11}\protect \) with \protect\( T\protect \)
for different concentrations of biphenyl in 8CB. The vertical lines denote the
transition temperatures \protect\( T_{AN}\protect \) and \protect\( T_{NI}\protect \)
for different concentrations of biphenyl in 8CB as shown in Table \ref{transition temp(table)}. }
\end{figure}

\begin{figure}
{\par\centering \resizebox*{0.9\textwidth}{!}{\includegraphics{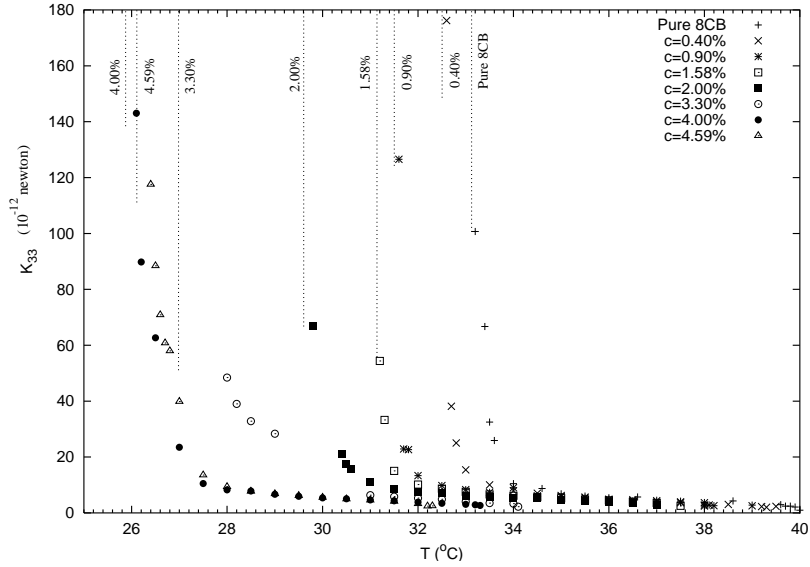}} \par}

\caption{\label{K33}Variation of \protect\( K_{33}\protect \) with \protect\( T\protect \)
for different concentrations of biphenyl in 8CB}
\end{figure}

The temperature and concentration dependence of the splay elastic constant \( K_{11} \)
is shown in Fig. \ref{k11}. The ratio \( \kappa  \) was seen to approach the
value 0 and stay around that value at and around transition but then it was
seen to increase again with increase in temperature. The plot of bend elastic
constant \( K_{33} \) with \( T \) is shown in Fig. \ref{K33} . 

The bend elastic constant \( K_{33} \) is seen to diverge as the smectic phase
is approached from the nematic side. The behavior was of the type \( \Delta K_{33}\propto t^{\nu } \),
where \( \Delta K_{33} \) is the difference \cite{morris86} between \( K_{33} \)
and its nematic part, \( t=(T/T_{AN}^{*}-1) \) and \( \nu  \) is the critical
exponent. We found that for all the samples studied \( \nu  \) is \( 1.0\pm 0.1 \)
which is in agreement with the exponent obtained by Morris et al \cite{morris86}
in pure 8CB. It must be recalled, however, that anisotropic scaling laws, predict
that \( \Delta K_{33} \) should vary as the correlation length \( \xi _{\Vert } \)
both above and below \( T_{AN} \) and this should result in an exponent \( \nu _{\Vert } \),
which from X-ray scattering experiments in many samples turn out to be \( \simeq 0.57-0.75 \)
\cite{deGennesch2}. But the temperature \( T_{AN}^{*} \) at which \( K_{33} \)
diverges is slightly higher by \( \sim 1^{0}K \) than \( T_{AN} \) for the
mixtures and for ease of eye, indicated by the vertical lines in Fig.\ref{K33}.
We are however unable to say if the weakly first order \( S_{A}-N \) transition
in pure 8CB approaches a second order transition in the mixtures and this needs
more detail studies. 

\begin{figure}
{\par\centering \resizebox*{0.9\textwidth}{!}{\includegraphics{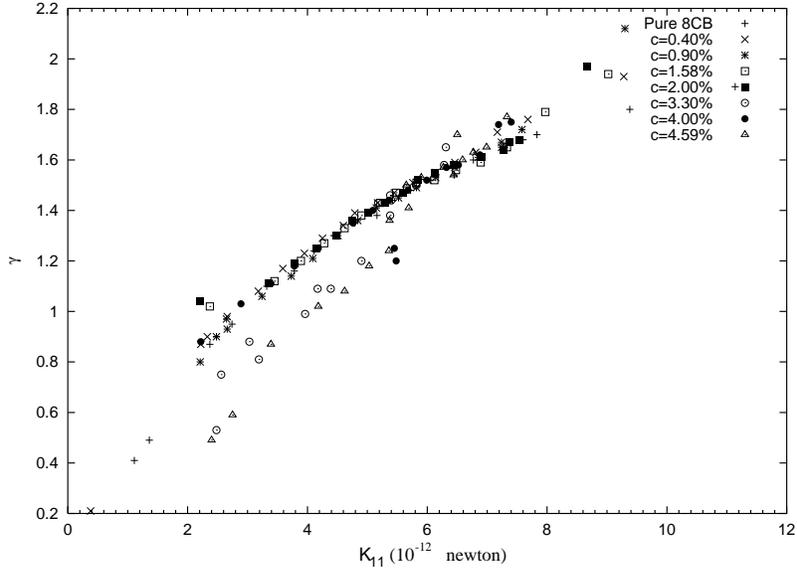}} \par}

\caption{\label{K11vsgamma}Variation of \protect\( K_{11}\protect \) with \protect\( \gamma \protect \)
for different concentrations of biphenyl in 8CB}
\end{figure}

On plotting \( K_{11} \) against \( \gamma  \) we find that the variation
for all the concentrations lie nearly on a universal curve, as shown in Fig.\ref{K11vsgamma}.
The deviations which are seen in the figure are mainly the points for the higher
concentration mixtures which lie in the vicinity of the two transitions. 

The abrupt decrease in \( \epsilon _{\Vert } \) as the smectic phase is approached,
seems to occur at relatively higher concentrations of biphenyl and is totally
absent in pure 8CB and in the mixtures of low impurity concentration. The optical
texture studies we have performed fail to reveal the existence of any smectic
phase other than smectic A in the mixtures at low temperatures. It may however
be conjectured that the increase in the proportion of biphenyl in the mixture,
though small, is leading to a type of ordering, cybotactic or otherwise, of
the 8CB molecules where the contribution of the permanent dipole moment of the
\( -CN \) group to \( \epsilon _{\Vert } \) is greatly reduced. 

The decrease in \( \epsilon _{\Vert } \) which suggests the existence of pre-transitional
smectic effects within the nematic phase occurs only for biphenyl concentrations
greater than a limiting value. It may be noted that the ratio \( T_{AN}/T_{NI} \)
changes from \( 0.9779 \) in pure 8CB to \( 0.9735 \) in the \( 4.59\% \)
biphenyl-8CB mixture. It may be pointed out that the decrease in this ratio,
though marginal, is believed to favor a second order \( S_{A}-N \) transition. 

\begin{figure}
{\par\centering \resizebox*{0.9\textwidth}{!}{\includegraphics{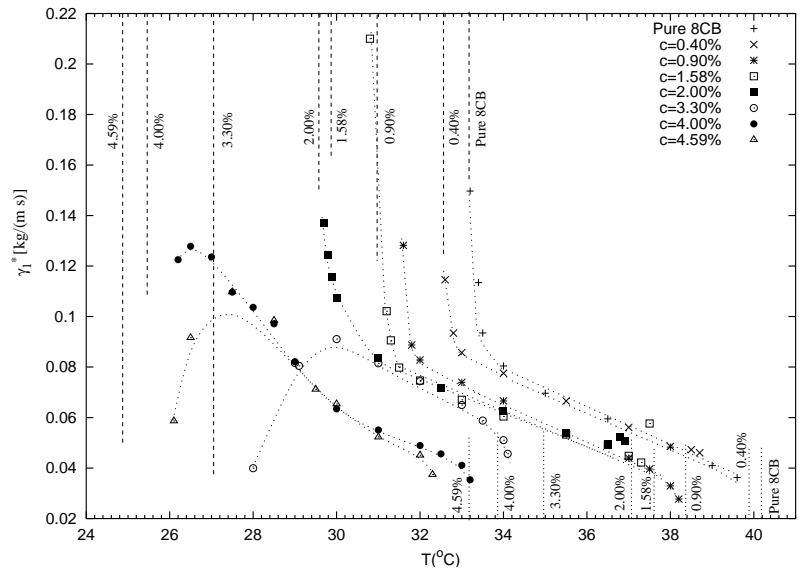}} \par}

\caption{\label{rotviscvsT}Variation of \protect\( \gamma _{1}^{*}\protect \) with
\protect\( T\protect \) for different concentrations of biphenyl in 8CB. The
vertical lines denote the transition temperatures \protect\( T_{AN}\protect \)
and \protect\( T_{NI}\protect \)}
\end{figure}

The temperature dependence of the rotational viscosity coefficient for the concentration
of biphenyl ranging from \( 0 \) to \( 4.59\% \) is shown in Fig.\ref{rotviscvsT}.
The \( T_{NI} \) (upper) is the temperature at which the field of view of the
polarizing microscope (with crossed polarizer and analyser) becomes completely
dark. In pure 8CB and for all the mixtures \( \gamma _{1}^{*} \) in the neighborhood
of \( T_{NI} \), shows the same feature in that it disappears at or before
\( T_{NI} \) (upper). However, as we approach \( T_{AN} \) from the higher
temperature side we find that for impurity concentrations upto \( 2.00\% \)
\( \gamma _{1}^{*} \) shows a sharp increase as in pure 8CB while for impurity
concentration of and higher than \( 3.30\% \), \( \gamma _{1}^{*} \) is seen
to decrease sharply. We are not aware of any systematic study involving te effect
of non-mesogenic solutes on the rotational viscosity coefficient of the nematics
in the neighborhood of nematic-smectic transition. The Miesowicz viscosities
of a mixture of 8OCB and 4TPB (exhibiting the nematic-smectic transition) were
recently investigated by Janik et al \cite{janik98} using a Miesowicz viscometer.

\begin{figure}
{\par\centering \resizebox*{0.9\textwidth}{!}{\rotatebox{270}{\includegraphics{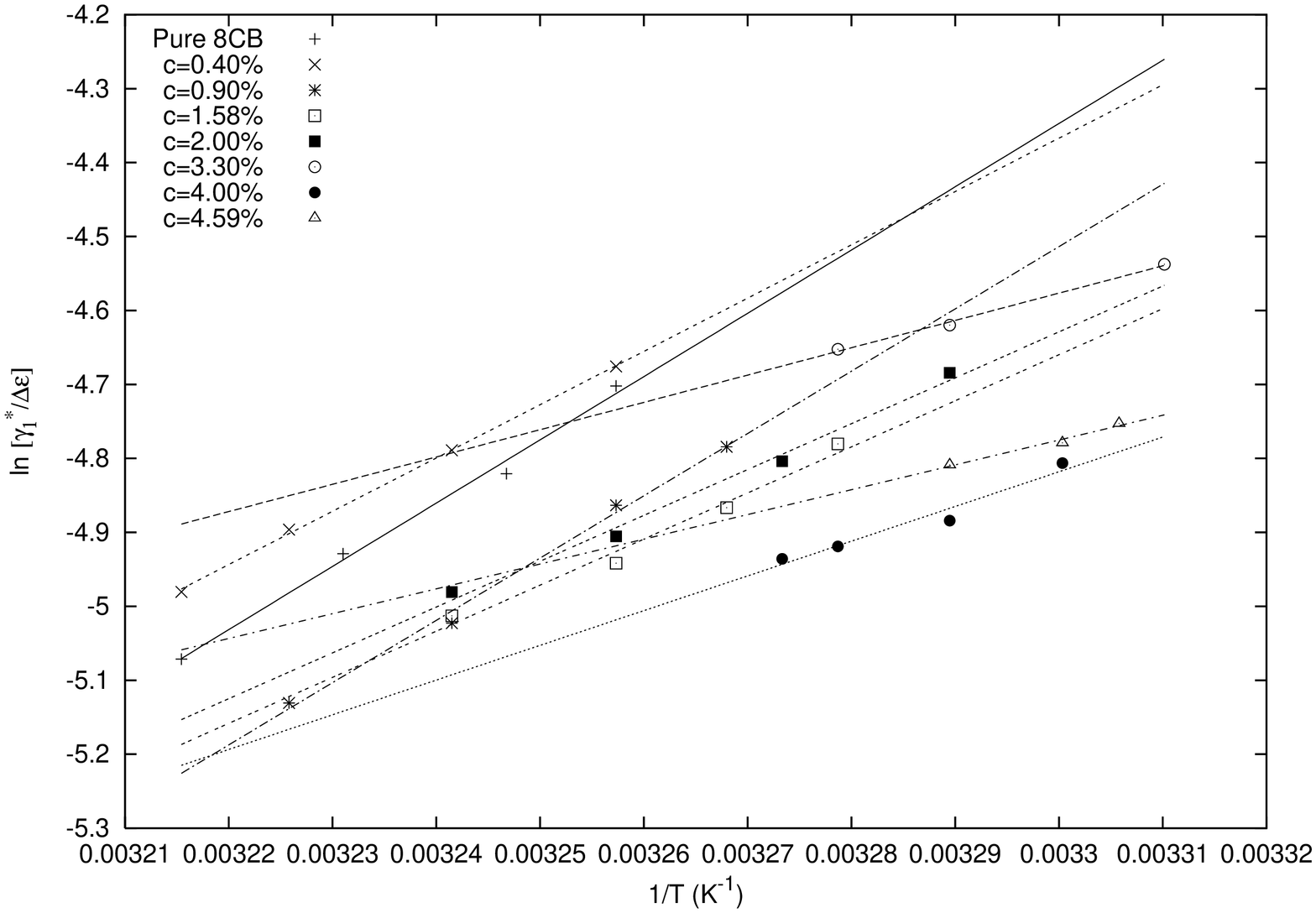}}} \par}

\caption{\label{orderparm}Variation of \protect\( \ln (\gamma _{1}^{*}/\Delta \epsilon )\protect \)
vs. \protect\( 1/T\protect \) for different concentrations of biphenyl in 8CB}
\end{figure}

Fig. \ref{orderparm} shows the plot of \( 1/T \) vs. \( \ln (\gamma _{1}^{*}/\Delta \epsilon ) \).
The dielectric anisotropy \( \Delta \epsilon  \) is roughly proportional to
the long range nematic order parameter. For pure as well as all mixtures the
plot of \( 1/T \) vs. \( \ln (\gamma _{1}^{*}/\Delta \epsilon ) \) gives a
straight line. The same sort of dependence has been obtained by Gasparoux and
Prost \cite{gasparoux71}, Heppke and Schneider \cite{heppke72} for MBBA and
by Prost et al \cite{prost76}, for a mixture of two isomers of \( p-methoxy-p'-butylazoxybenzene \).
Close to \( T_{NI} \) and \( T_{AN} \) there are deviations and those points
have not been shown in the figure. From the slope of these plots the activation
energy for diffusion \cite{dejeu} was calculated and has been shown in Table
\ref{activation energy}.

\section{CONCLUSIONS}

To summarize it can be stated that the presence of the rigid, non-polar and
non-mesogenic solute biphenyl in 8CB results in a change in the dielectric anisotropy
\( \Delta \epsilon  \), the splay elastic constant \( K_{11} \) and the effective
rotational viscosity coefficient \( \gamma _{1}^{*} \). In the neighborhood
of the nematic-smectic A transition for the concentration of biphenyl \( \geq 3.3\% \)
the changes in these quantities have been found to be anomalous in that, instead
of the usual rapid increase observed on cooling, \( \Delta \epsilon  \), \( K_{11} \)
and \( \gamma _{1}^{*} \) sharply decrease. This may perhaps be attributed
to the presence of pretransitional effects within the nematic phase. From optical
texture studies we could not identify any new smectic phase in the mixtures
other than what is found in 8CB. The bend elastic constant \( K_{33} \) in
all the mixtures was found to diverge as the smectic phase was approached and
the critical exponent was found to be the same as that observed for pure 8CB,
although the temperatures at which the divergences occur were slightly higher
than the respective values of \( T_{AN} \). More elaborate studies involving
DSC, X-ray diffraction and refractive index measurements (to obtain the long
range nematic order parameter) may perhaps reveal more about what is actually
happening in the mixtures near the smectic A-nematic transition particularly
for those containing relatively higher biphenyl concentration.

\begin{table}
{\centering \begin{tabular}{|c|c|c|c|}
\hline 
Concentration (\%)&
\( T_{AN} \) &
\( T_{NI} \) (lower)&
\( T_{NI} \) (upper)\\
\hline 
\hline 
0.00&
33.2&
40.1&
40.2\\
\hline 
0.40&
32.8&
39.8&
39.9\\
\hline 
0.90&
31.0&
38.1&
38.4\\
\hline 
1.58&
29.9&
37.2&
37.6\\
\hline 
2.00&
29.6&
36.5&
37.1\\
\hline 
3.30&
27.1&
34.3&
35.0\\
\hline 
4.00&
25.5&
33.1&
33.9\\
\hline 
4.59&
24.9&
32.2&
33.2\\
\hline 
6.40&
-&
28.2&
29.8\\
\hline 
\end{tabular}\par}

\caption{\label{transition temp(table)}Variation of phase transition temperatures \protect\( T_{AN}\protect \)
and \protect\( T_{NI}\protect \) with concentration of biphenyl in 8CB}
\end{table}

\begin{table}
{\centering \begin{tabular}{|c|c|}
\hline 
Concentration (\%)&
Activation Energy (eV)\\
\hline 
\hline 
0.00&
0.74\\
\hline 
0.40&
0.62\\
\hline 
0.90&
0.73\\
\hline 
1.58&
0.54\\
\hline 
2.00&
0.53\\
\hline 
3.30&
0.32\\
\hline 
4.00&
0.40\\
\hline 
4.59&
0.29\\
\hline 
\end{tabular}\par}

\caption{\label{activation energy}Variation of activation energies with concentration
of biphenyl in 8CB}
\end{table}

\section*{ACKNOWLEDGEMENTS}

The authors acknowledge valuable discussions with Dr. A. DasGupta, Dr. P.K.
Mukherjee and Dr. K. Mukhopadhyay. The research was supported by DST (grant
no. SP/S2/M-20/95). SDG acknowledges the award of a fellowship.

\vspace{0.5001cm}

\begin{thebibliography}{10}

\bibitem{martire76}
D.~E. Martire et~al.
\newblock {\em J. Chem. Phys.}, 64:1456, 1976.

\bibitem{peterson74}
H.~Peterson and D.~E. Martire.
\newblock {\em Mol. Cryst. Liq. Cryst.}, 25:89, 1974.

\bibitem{kroberg76}
B.~Kroberg, D.~Gilson, and D.~Patterson.
\newblock {\em J. Chem. Soc., Faraday II}, 72:1673, 1976.

\bibitem{cotter69}
M.~Cotter and D.~E. Martire.
\newblock {\em Mol. Cryst. Liq. Cryst.}, 7:295, 1969.

\bibitem{cotter76}
M.~Cotter.
\newblock {\em Mol. Cryst. Liq. Cryst.}, 35:33, 1976.

\bibitem{maier59}
W.~Maier and A.~Saupe.
\newblock {\em Z. Naturforsch}, 14a:882, 1959.

\bibitem{mukherjee97}
P.~K. Mukherjee.
\newblock {\em Liq. Cryst.}, 22:239, 1997.

\bibitem{deGennesch2}
P.~G. {de Gennes} and J.~Prost.
\newblock {\em The Physics of Liquid Crystals}.
\newblock Oxford Science Publications, 1993.

\bibitem{karat77}
P.~P. Karat and N.~V. Madhusudana.
\newblock {\em Mol. Cryst. Liq. Cryst.}, 40:239, 1977.

\bibitem{bradshaw85}
M.~Bradshaw, E.~Raynes, J.~Bunning, and T.~E. Faber.
\newblock {\em J. Physique}, 46:1513, 1985.

\bibitem{morris86}
S.~W. Morris, P.~{Palffy-Muhoray}, and D.~Balzarini.
\newblock {\em Mol. Cryst. Liq. Cryst.}, 139:263, 1986.

\bibitem{pinku94}
P.~Chattopadhyay and S.~K. Roy.
\newblock {\em Mol. Cryst. Liq. Cryst.}, 89:257, 1994.

\bibitem{gruler}
H.~Gruler, T.~J. Scheffer, and G.~Meier.
\newblock {\em Z. Naturforsch}, 72a:966, 1972.

\bibitem{meyerhofer}
D.~Meyerhofer.
\newblock {\em J. Appl. Phys.}, 46:5084, 1975.

\bibitem{dejeu}
W.~H. {de Jeu}.
\newblock {\em Physical Properties of Liquid Crystalline Materials}, volume~1.
\newblock Gordon and Breach Science Publishers.

\bibitem{pieranski73}
P.~Pieranski, F.~Brochard, and E.~Guyon.
\newblock {\em J.Phys}, 34:35, 1973.

\bibitem{rosenblatt82}
C.~Rosenblatt.
\newblock {\em Phys. Rev. A}, 25:1239, 1982.

\bibitem{deJeu74}
W.~H. de~Jeu, W.~J.~A. Goosens, and P.~Bordewijk.
\newblock {\em J. Chem. Phys.}, 61:1985, 1974.

\bibitem{janik98}
J.~Janik, J.~K. Moscicki, K.~Czuprynski, and R.~Dabrowski.
\newblock {\em Phys. Rev. E}, 58:3251, 1998.

\bibitem{gasparoux71}
H.~Gasparoux and J.~Prost.
\newblock {\em J. Phys. Paris}, 32:953, 1971.

\bibitem{heppke72}
G.~Heppke and F.~Schneider.
\newblock {\em Z. Naturforsch}, 27:976, 1972.

\bibitem{prost76}
J.~Prost, G.~Sigand, and B.~Regaya.
\newblock {\em J. Phys. Lett., Paris}, 37:L341, 1976.

\end{thebibliography}
\end{document}